# Reaction Kinetics in the Production of Pd Nanoparticles in Reverse Microemulsions. Effect on Particle Size.

Jhon Sanchez-Morales[1], Miguel D. Sánchez[1] and Hernán A. Ritacco[(*)1].

[1] Instituto de Física del Sur (IFISUR), Departamento de Física, Universidad Nacional del Sur (UNS), CONICET, Av. L. N. Alem 1253, *B8000CPB - Bahía Blanca, Argentina*.

(*) Corresponding author: hernan.ritacco@uns.edu.ar

**ABSTRACT**

In the synthesis of metallic nanoparticles in microemulsions, we hypothesized that particle size is mainly controlled by the reaction rate. Thus, the changes observed on the particle sizes as reaction conditions, such as concentrations, temperature, type of surfactant used, etc., are varied should not be correlated directly to the modification of those conditions but indirectly to the changes they produce on the reaction rates.

By means of time resolved UV-vis spectroscopy, we measured the reaction rates in the production of Pd nanoparticles inside microemulsions at different reactant concentrations, keeping all the other parameters constant. The measured reaction rates were then correlated with the particle sizes measured by transmission electron microscopy (TEM).

We found that nanoparticle size increases linearly as the reaction rates increases, independently of the actual reactant concentrations. We proposed that the kinetics is controlled mainly by the diffusion of the reducing agent through the surfactant monolayer covering the microemulsion membrane. With this model, we predicted that particle size should depend indirectly, via the reaction kinetics, on the micelle radius ($\sim r^{-3}$), the water volume ($\sim v_w^3$) and the total microemulsion volume ($\sim v_T^{-3}$), and temperature (Arrhenius). Some of these predictions were explored in this article.

## INTRODUCTION

Microemulsions[1] are thermodynamically stable dispersions of immiscible liquids, in general water and an oil, that are stabilized by some chemical agent, in general a surfactant[2]. Surfactants are amphiphilic molecules containing different chemical groups on their structures, some of them with affinity for one of the liquids in the dispersion, and some others with affinity for the other liquid. Because of this, they place themselves spontaneously on the interface between the liquids, exposing their chemical groups in the direction of the liquid for which they have affinities, reducing the surface tension, and stabilizing in this way the dispersion. The synthesis of nanoparticles in reverse microemulsions was proposed in the 1980s as a method for obtaining monodisperse particles with perfectly controlled sizes. The term *reverse* or *water-in-oil* (W/O) microemulsions refers to droplets of water dispersed in a continuous phase of an oil, in opposition to *direct* or *oil-in-water* (O/W) microemulsion that refers to droplets of oil dispersed in water[1–3].

Since its formulation[4], the synthesis of nanoparticles via microemulsions[5] has become a widely used technique [6] for the preparation of a diversity of nanoparticles. There are several reviews on the subject[7–12]; those by Tovstun et al.[6] and Ganguli et al. [12] are the most recent ones, both very comprehensive and easy to read. Among the particles that can be produced by this method, a great variety of precursors have been used to obtain metallic nanoparticles including nickel[13], gold [14], silver[15], platinum [16], and palladium [17]. We are particularly interested in the synthesis of Pd nanoparticles due to their physicochemical properties and the possibility of using them for catalytic purposes[18–20]. These properties are strongly related to the size and structure of the obtained nanoparticles; hence, studies on growth control and size distribution have received a great deal of attention in recent years[21–25].

Although reverse microemulsions as nanoreactors have been used successfully to produce a number of nanostructures of different sizes and shapes[12,26], the role played by microemulsions in tailoring those structures is still unresolved. In the early days of the method, it was considered that the size of the particles obtained depended mainly on the size of the droplets in the microemulsion, the water pool acting as a template in defining the final size and the structure of the nanoparticle. However, it is not always possible to find a direct correlation between the size of the aqueous core of the droplets and the size of the synthesized nanoparticle[11]. Other investigations have revealed that the control of size is a complex phenomenon, which involves mechanisms of chemical kinetics and mass transfer between the micelles[27]. These mechanisms and, therefore, the final size of the nanoparticles would be affected by other control variables, such as the concentration of the reagents in the drops[28], the amount of water[23], the flexibility or rigidity of the surfactant film[29], the capability of the surfactants/co-surfactants in protecting the particles against aggregation and coarsening, and even the autocatalytic growth processes[30]. Additionally, an important number of published works have experimentally and theoretically studied the effects of micellar exchange with respect to the kinetics of the formation and growth of nanoparticles[31–34]. However, the literature on the experimental base is still scarce and the effect of intermicellar dynamics and control variables on the size of the particles is far from being understood. It is quite difficult to develop some general ideas about the synthesis of inorganic nanoparticles in reverse microemulsions from the "chaotic" amount of experimental results because of the huge number of parameters that influences the synthesis results[6,21,33,35,36]. There is a lack of systematic studies on the impact of each parameter on the nanoparticle formation and properties in well controlled systems, particularly as regards experiments on the reaction kinetics inside microemulsions and its effect on particle size and morphology[36]. In most of the published works about the production of metallic nanoparticles in microemulsions, the kinetics of particle growths is commonly measured following the temporal evolution of the surface Plasmon Resonance in UV-vis spectroscopy. However, these dynamics include, if present, the processes of coalescence and coarsening of the particles superimposed to the kinetics of the chemical reaction. In this article, we study systematically the reaction kinetics of metallic nanoparticles of Palladium in reverse microemulsions. Palladium metallic nanoparticles do not present a Surface Plasmon resonance in the UV-vis region, but the palladium salt precursor used, $PdCl_4^{-2}$, has an adsorption peak at 425 nm, which allows us to follow the reaction kinetics without interferences of coalescence and coarsening. We focused our attention on the relationship between the initial reaction rates and the final particle size. We changed the reaction rates by changing the reactant concentrations and temperature. A simple model is proposed in order to rationalize the experimental findings. This model predicts dependences of the reaction rates on the number and sizes of the microemulsions droplets, the total system, and water volumes and temperature. Some of these predictions were verified in the present work, some were observed on data published by other researchers, and some other ones will be explored in future work. We found a linear correlation between the initial reaction rates and the final particle size independently of the actual concentrations or temperature. In other words, for this particular system, the nanoparticle size depends indirectly on reactant concentrations and temperature via the reaction rate.

## MATERIALS AND METHODS

*2.1. Chemicals, main reaction and microemulsion preparations*

Palladium(II) chloride ($PdCl_2$), sodium borohydride ($NaBH_4$), (n-Dodecyl) trimethylammonium bromide (DTAB), hydrochloric acid (HCl), n-octanol ($C_8H_{18}O$), and Benzene ($C_6H_6$) were used in the formulation of the microemulsions. All reagents were of analytical grade, and they were used directly without further purification. The water used to prepare the solutions was from Mili-Q® ultrapure water purification system.

Aqueous solutions of Pd (II) (~ $10^{-3}$ mol dm$^{-3}$) that were obtained by dissolving $PdCl_2$(s) in $HCl/H_2O$ (6 mM) were prepared. We used $NaBH_4$ solutions in water at different concentrations ranging from 30 to 120 mM as a reducing agent. Two identical microemulsions were prepared with reverse micelles. Precisely measured volumes (~ $10^{-6}$ dm$^{-3}$) of the precursor solutions were injected into 5 ml of a benzene/DTAB/n-octanol mixture, [DTAB]=0.1M, [octanol]=0.7 M. The molar ratio of water to DTAB in each microemulsion was $W_0$=*15*.

The nanoparticles were synthesized by rapidly mixing equal volumes (1 cm$^3$) of the microemulsions, with one volume containing palladium ions and the other one containing adequate amounts of NaBH$_4$ as a reducing agent. The reaction kinetics was measured by following the reduction of the absorbance peaks at a wavelength of 340 nm by Ultraviolet-visible light spectroscopy, UV-vis. This peak is characteristic of the complex [PdCl$_4$]$^{-2}$, and disappears as the reaction advances to produce metallic Pd particles.

The mechanism of the reaction should be considered when discussing the kinetics of particle production in the following sections. In particular, we want to stress that the chemical reaction involved in the production of Pd particles consists of two reactions: the hydrolysis of the reducing agent that ends in a release of gaseous hydrogen [37],

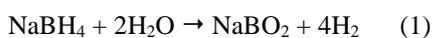
$$NaBH_4 + 2H_2O \rightarrow NaBO_2 + 4H_2 \quad (1)$$

and the reduction of palladium(II) ions (in HCl (aq)) by gaseous hydrogen[38]

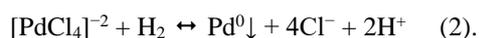
$$[PdCl_4]^{-2} + H_2 \leftrightarrow Pd^0\downarrow + 4Cl^- + 2H^+ \quad (2).$$

Due to the fast hydrolysis and the production of H$_2$, all experiments of the reaction kinetics were performed with NaBH$_4$ solutions freshly prepared: no more than 10 minutes elapsed between the preparation of the microemulsion with NaBH$_4$ and its use in the reaction kinetics experiments. It is relevant to mention here that after about 2 hs, the NaBH$_4$ aqueous solutions lost their capacity of acting as a reduction agent. This is a clear indication of the production of H$_2$ in the hydrolysis.

*2.3. Characterization Methods*

The size of the micelles in the microemulsion was determined with the dynamic light scattering (DLS) technique by using a multi-angle Malvern auto-sizer 4700 system equipped with an OBIS laser (λ=514nm, 20 mW). Measurements of samples that were identical to the microemulsions used in the synthesis were made in order to verify the stability of the microemulsions.

As previously mentioned, the reaction kinetics was followed by time resolver UV-vis spectrometry after mixing two identical microemulsions: one containing the Pd ions and the other one containing the reduction agent. All the measurements were conducted in an Ocean Optics USB2000 miniature spectrophotometer with a measuring wavelength range of 200-800 nm. A quartz cuvette with an optical path of 1 cm was used.

The particle size distribution was performed manually from the digital analysis of the TEM images. A few drops of the "micellar" solution containing the palladium solid material were taken and deposited directly on carbon-coated copper grids specifically designed for microscopy. The samples were allowed to dry for 12 hours, and they were then analyzed using a JEOL-100CX II transmission electron microscope (TEM) operating at 100 kV and a FEI Talos™ F200A high-resolution transmission electron microscope (HR-TEM) operating al 200 kV. For each sample, the average size and standard deviation were determined from a minimum of 250 particles from different parts of the grid.

## RESULTS

**Effect of precursors concentration in microemulsión sizes.**
Figure SI-x shows the hydrodynamic diameter distribution of microemulsion droplets as measured by Dynamic Light Scattering. We performed the measurements in microemulsions containing different amounts of PdCl$_2$ and also NaBH$_4$ in order to confirm that no changes occur in the microemulsion droplet size. The apparent hydrodynamic diameter of the micelles was about 10 nm, independently of the chemical concentrations. We also measured the droplet size after reaction. No appreciable variations in the average diameter of the micelles were observed after the reaction was completed.

Likewise, no appreciable variations in time were observed for the size distribution of the individual microemulsions containing the reagents. The measurements were made every 24 h during five consecutive days.

**Reaction kinetics of the formation of metallic Pd nanoparticles.**

As previously mentioned, the reaction kinetics in the formation of Pd particles was followed by UV-vis spectrophotometry. In the typical spectra shown in Fig. 1a, two absorption peaks were identified at wavelengths 280 nm and 340 nm, respectively. Due to the medium conditions (pH=2.5), these peaks have been assigned to the UV-vis peaks that are only characteristic of the ions PdCl$_4$$^{-2}$ [39]. The same peaks appear in aqueous solutions of PdCl$_2$ in the

presence of HCl at pH = 2.5, but shifted to 325 and 425 nm, respectively (see Figure SI-xx). The shift is a consequence of the confined environment of the microemulsions. Because the observed transition moves to shorter wavelengths (higher energies), it is probable that the negatively charged complex $PdCl_4^{-2}$ is stabilized by the oppositely charged interface covered by DTAB positively charged ions. Note that the position of the peak does not change as the concentration of the Palladium precursor is changed in the microemulsion (see Figure SI-XX), assuring that the measured kinetics is not affected by a spurious movement of the position of the peak. Additionally, it was found that the microemulsions containing the $NaBH_4$ as a reducing agent do not absorb electromagnetic radiation in the range of 275 to 580 nm. Therefore, the progress of the reaction can be followed by the variation of the absorbance of the palladium(II) ions during the reduction at the mentioned wavelength. This is shown in Figure 1b, where the absorbance was transformed in $Pd^{2+}$ concentrations by a calibration curve (shown in Figures SI-XX of the supporting information document). In all cases, a decrease in the UV-vis signal was observed up to a stable point associated with the end of the reaction. Given the amounts of the reducing agent, it can be assumed that the conversion of Pd(II) is close to 100% at all the concentrations explored.

In order to determine the rate law of the reduction reaction, experiments were performed under different conditions of concentration of the reagents.

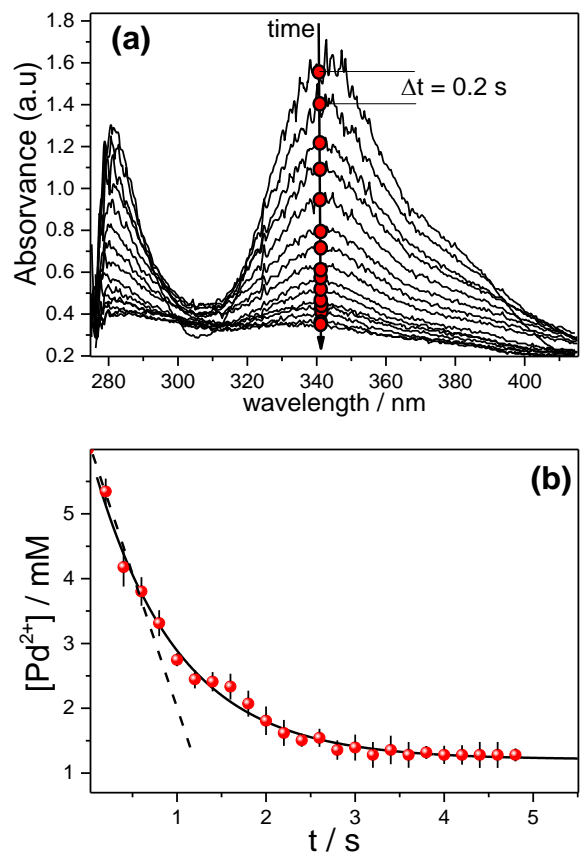

Figure 1: (a) UV-vis spectra measured at time intervals of 0.2 s during the reaction in the microemulsion. The peak at 340 nm is followed in order to determine the rate of reaction plotted in (b). The results correspond to the reaction with initial concentrations of $[PdCl_2]_0$= 6 mM; $[NaBH_4]_0$= 90 mM; $W_0$=15.

From the curves like that the one shown in Figure 1b, we obtained from the slopes the initial rates of the reactions, $v_0 = -\frac{d[Pd^{2+}]}{dt}$, as a function of the initial concentration of the reduction agent, $[NaBH_4]_0$, and for different concentrations of the precursor, $[PdCl_2]$. The results are presented in Figure 2.

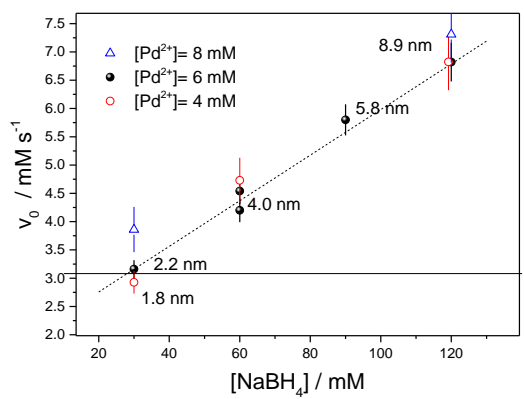

Figure 2: Initial reaction rate as a function of concentration of the reduction agent for different concentrations of the Palladium salt. The mean diameter of the nanoparticles obtained after the reaction are inserted in the figure.

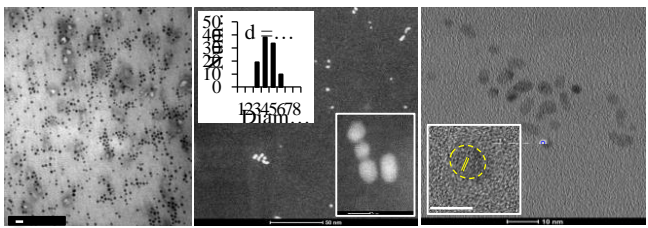

*Figure 3: Analysis of the nanoparticles by high-resolution transmission electron microscopy (HR-TEM). (a) TEM micrograph of nanoparticles obtained with [NaBH₄] = 60 mM and [Pd²⁺] = 6 mM. (b) Analysis using dark-field TEM and determination of size. (c) HR-TEM image. A typical Pd nanoparticle is shown.*

It can be observed that an increase in the reducing agent concentration leads to an increase in the reaction rate for the three concentrations of palladium studied, $[Pd^{2+}]$ = 8, 6 and 4 mM, being the reaction rate increment linear with the concentration of the reducing agent and with the same slope for the three concentrations of palladium. The reaction rates seem to be independent of the palladium concentration, within the errors and for 4 <= $[Pd^{2+}]$ <= 8. This is also clearly seen in Figure 3, where we present results of the initial reaction rates as a function of palladium concentration keeping the concentration of the reducing agent constant. The reaction rate does not change when the precursor, the palladium salt, concentrations are changed to a constant concentration of NaBH₄.

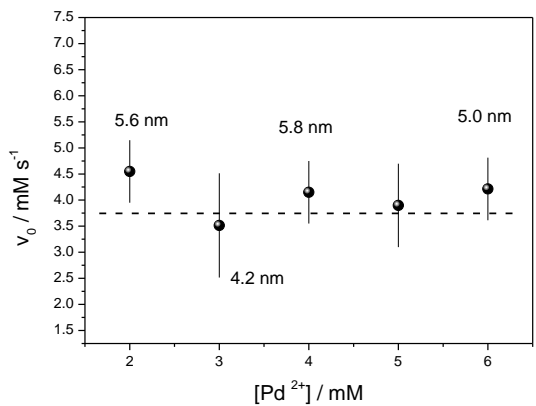

*Figure 4 : Initial reaction rate as a function of the precursor concentration, PdCl₂. The initial concentration of the reducing agent, NaBH₄, is fixed at 60 mM. The mean diameter of the obtained metallic Palladium nanoparticles is inserted on the figure.*

**Particle size, size distribution and morphology.**

Once the equilibrium condition (constant absorbance value, see Fig. 1b) was reached, we collected samples of each microemulsion containing the synthetized palladium nanoparticles. The samples were analyzed by TEM in order to characterize them and to determine their size. Figure 4 shows images of the nanoparticles that were obtained for initial concentrations of [NaBH₄] = 60 mM and [Pd²⁺] = 6 mM. The sizes that were confirmed by dark-field TEM micrographs to be an average diameter of 4.4 nm show monodisperse Pd nanoparticles. The shape of the particles was approximately spherical. Such small spherical and monodisperse Pd particles are quite difficult to obtain [40]. This morphology is typical of palladium metal particles with face-centered cubic (fcc) symmetry. In the images, the lattice fringes of the particles that would be related to a crystalline structure were clearly observed. The interplanar distance of 2.24 Å would correspond to the planes (111) of the face-centered cubic (fcc) structure of the metallic palladium.

Figure 5 shows representative images and the size distribution of the palladium nanoparticles that were synthesized in microemulsion at different reagents concentrations. From the results, it can be seen that the average particle size increases as the amount of the reducing agent increases in relation to palladium. The apparent average diameter of the synthesized nanoparticles varies between 2 and 9 nm, with a standard deviation that increases as the concentration of NaBH₄ increases.

All the measured sizes at the different reaction conditions were also indicated in the previous Figures 2 and 3. Note that, in all cases, the average diameter of the nanoparticles is less than the apparent hydrodynamic diameter of the micelles in the microemulsion (~10 nm), as it was measured by Dynamic Light Scattering.

**DISCUSSION.**

On empirical bases, the initial reduction rate of the Pd(II) ions can be written as:

$$v_0 = k_{obs} \cdot [Pd^{2+}]_0^n \cdot [NaBH_4]_0^m \qquad (3),$$

where $v_0$ is the initial reaction rate, $k_{obs}$ is the observed reaction rate constant, $[Pd^{2+}]_0$ and $[NaBH_4]_0$ are the initial concentrations of the reactants, $n$ and $m$ are the reaction orders with respect to the precursor and the reducing agent, respectively.

From the previous results (on the basis of more than a

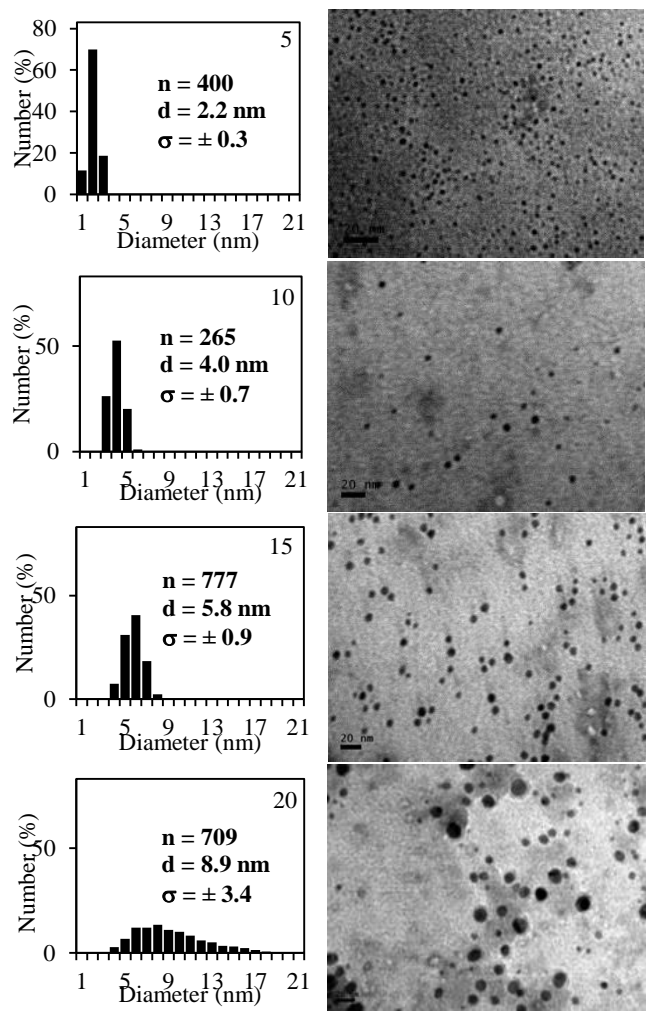

Figure 5: TEM micrographs and size distribution of the Pd nanoparticles obtained in microemulsion. Conditions from top to bottom, [NaBH$_4$] = 30, 60, 90 and 120 mM. In all cases [Pd$^{2+}$]=6 mM Room temperature. Scale: 20 nm.

hundred experiments), and for [Pd$^{2+}$] ≥4 mM, we found $n=0$ and $m=1$. Thus, the reaction is independent of the palladium concentration and first order in the reducing agent (and globally). Under these conditions and from the slope of Figure 2 (considering all the performed experiments), the rate constant gives, $k_{obs}$ = (0.040±0.002) s$^{-1}$ (note that if hydrogen concentration, [H$_2$] had been used instead of [NaBH$_4$] in Figure 2, and considering the stoichiometry of the hydrolysis -see material section-, the kinetic constant would have given $k_{obs}/4 = 0.01$ s$^{-1}$).

In order to propose a possible reaction mechanism, let us stress some facts. The experimental observation that the reaction rate is first order for the reducing agent and zero order for the Pd salt could indicate that the reaction is controlled by the diffusion of the reduction agent, i.e. H$_2$. Because the ionic species, PdCl$_4^{-2}$, involved in the reaction could hardly be transferred through the oil phase[41], let us assume then that they remain in the water pool of the microemulsion. Thus, the unique way of transferring PdCl$_4^{-2}$ from one water pool to another is by collision and fusion of the micelles[42–47]. The situation for H$_2$ molecules is quite different; the solubility of H$_2$ is four times larger in benzene than in water. We already mentioned that the reducing properties of the microemulsion containing NaBH$_4$ disappear after 2 hs, which seems to indicate the transfer of H$_2$ out of the water pool of the microemulsion. Thus, the evidence indicates that H$_2$ can diffuse from one water pool to another through the continuous phase (oil). Now, recall that the microemulsion is stabilized with the cationic surfactant DTAB; the DTA+ ion can then interact with the oppositely charged ion PdCl$_4^{-2}$ at the droplet interface. This possibility is supported by the experimental observation of the shifting of the absorption peak in the UV-vis spectroscopy from 425 to 340 nm when the ion PdCl$_4^{-2}$ is placed into the microemulsion (see material section). Thus, we could think that the reaction occurs at the micelle interface. Based on all the previous comments, we propose the following reaction mechanism, schematized in Figure 6.

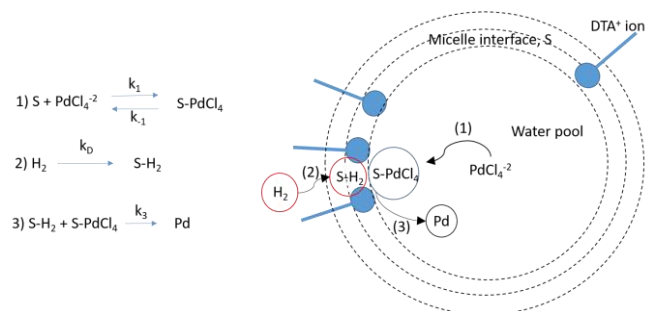

Figure 6: Proposed reaction mechanism. The reaction would be mediated by the micelle interface in a three-step mechanism, being the diffusion of H$_2$ the controlling step in the reaction kinetics.

We propose a three-step mechanism where the reaction is mediated by the micelle interface. The first step of the mechanism is an equilibrium with direct and inverse kinetic constants $k_1$ and $k_{-1}$; the DTA+ ions of the surfactant in the micelles interface provide sites (S) for the condensation of oppositely charged precursor PdCl$_4^{-2}$, given the interface complex S-PdCl$_4$. The second step is the diffusion of H$_2$ through the walls of the micelles into the interfacial region, S. A hydrogen molecule in that region is represented by S-H$_2$, being the kinetic constant of the process $k_D$. The third step is the reaction of the precursor and the reducing agent in the interfacial region. The kinetic constant in this case is $k_3$. With this mechanism, the reaction rate is read (see SI),

$$v = \frac{k_D k_1 [H_2][S]_0 [PdCl_4^{-2}]}{k_{-1} + k_D[H_2] + k_1 [PdCl_4^{-2}]} \quad (4),$$

where $[S]_0$ is the total concentration of interfacial sites on the micelle interface (probably related with the number of DTAB ions on the micelle interface). If the concentration of the precursor is large enough so that it saturates all the interfacial sites, S, and $k_D$ and $k_{-1}$ are smaller than $k_1[PdCl_4^{-2}]$, then the reaction rate becomes independent of $[PdCl_4^{-2}]$ and is controlled by the diffusion of the reducing agent,

$$v \sim k_D [S]_0 [H_2] \quad (5).$$

Note that the reaction rate law is of pseudo-first order in $H_2$ and zero order in Pd precursor, as it was observed experimentally. The kinetic constant $k_D$ corresponds to the diffusion of $H_2$, in the second step of the reaction mechanism into the micelles containing the Pd precursor; it can be expressed in terms of the diffusion coefficient,

$$v_2 \sim J\,A = -D_{eff} A \frac{d[H_2]}{dx} \sim D_{eff} A \frac{\Delta[H_2]}{\Delta x} = D_{eff} \frac{A_m (N/2)}{l}[H_2]$$
$$(6),$$

where we have used the one-dimensional Fick's law in order to write the reactant diffusion flux, $J$ (moles per unit of area and time); A is the total area of the micelles per unit volume (A $\sim$ (N/2). $A_m$; $A_m = 4\pi r^2$ being r the micelle radius and N the number of micelles per total volume) and $l$ the diffusion distance. Thus, the kinetic constant is read,

$$k_D = D_{eff} \frac{A_m (N/2)}{l} \quad (7).$$

Equations 7 predicts that the kinetic constant $k_D$ depends on the size of the micelles in the microemulsion via the area, $A_m$, the concentration of micelles, N, and the distance between them, $l$. Note that N is the number of micelles per unit volume, $N \approx \frac{V_w}{(4/3 \pi r^3)} \frac{1}{V_T}$, being $V_w$ the volume of water added, r the radius of the micelles, and $V_T$ the total volume of the microemulsion. Because $l$ can be estimated as, $l \sim \frac{1}{N\pi r^2}$, then the kinetic constant should scale as,

$$k_D \sim \frac{9}{8} D_{eff} \left(\frac{V_w}{V_T}\right)^2 \frac{1}{r^2} \quad (8).$$

Introducing this equation in Equation 5, we obtain for the initial reaction rate

$$v_0 \sim \frac{9}{8} D_{eff} \left(\frac{V_w}{V_T}\right)^2 \frac{1}{r^2} [S]_0 [H_2] = k_{obs}[H_2] \quad (9)$$

being

$$k_{obs} = \frac{9}{8} D_{eff} \left(\frac{V_w}{V_T}\right)^2 \frac{1}{r^2} [S]_0 \quad (10).$$

We can now make some estimations: the total volume, $V_T$, in our systems was always 2 ml, the water volume, $V_W$ = 50 μl, and the micelles radius, $r$ = 5 nm. In order to calculate $k_D$ (Eq. 8), we just need the effective diffusion coefficient, $D_{eff}$. The diffusion constant for the hydrogen in benzene liquid is D $\sim 10^{-9}$ m$^2$s$^{-1}$; however, $H_2$ molecules have to additionally diffuse through two micelle membranes formed by surfactant monolayers. These monolayers can exert a high resistance to mass transfer[48]. Moreover, the interfacial region exerting the resistance to the diffusion of material is not limited to the actual surfactant film thickness because the presence of the surfactant modifies the properties of the liquid in the vicinity of the interfaces at longer distances. It was found that the effective diffusion coefficients for gases in the interfacial region can be three to four orders of magnitude smaller than in bulk liquids[48]. A similar behavior is found in a similar physical phenomenon: the coarsening in liquid foams[49], where gas diffuse through two surfactant monolayers separating adjacent bubbles. Thus, a reasonably estimation for the effective diffusion coefficient should be $D_{eff}$ = 10$^{-12}$ - 10$^{-13}$ m$^2$s$^{-1}$. Using all these values in Equation 8 results in $k_D \sim$ 2.3 to 23 s$^{-1}$. From the observed kinetic constant, $k_{obs}$, the concentration of interfacial sites can be estimated; for $k_D$=2.3s$^{-1}$, we obtain $[S]_0 \sim$ 4.510$^{-3}$ M and the number of sites per micelle is about $\sim$110 sites/micelle. Just for comparison, it is worth mentioning that the number of DTAB molecules per micelle in the microemulsions is about 1200; then, about 10% of them should be acting as interfacial binding sites.

Equation (9) predicts changes of the reaction rate with the size of the microemulsion droplet. The concentration of interfacial sites can be considered as proportional to the total interfacial area, which is the product of the number of micelles by the

micelle area, $[S]_0 \sim NA_m$. Using the previous expressions for N and $A_m$ into Equation (9) yields

$$v_0 \sim D_{eff} \left(\frac{V_w}{V_T}\right)^3 \frac{1}{r^3} [H_2] \qquad (11).$$

This equation predicts a dependence of the reaction rate on water and total volume, $(V_w/V_T)^3$. It is worth mentioning that the effect of microemulsion water and total volume, $V_w$ and $V_T$, on the size of particles obtained by reaction in microemulsions was recently reported[23].

Note that when $v_0$ scales as, $v_0 \sim r^{-3}$, the initial reaction rate should diminish with the micelle radius to the power 3, if keeping all the other parameters constant. This prediction is interesting; if we accept that the particle size and the size distribution are affected by the reaction rate, then Eq. (9) predicts changes on the particle size as the microemulsion size (r), changes. However, this is not due to the modification of the "nanoreactor" size acting as a template, but indirectly due to the reaction kinetics. In this way, the particle size is not a consequence of the reaction taking place in the confined water pool of the microemulsion ; then, it is not limited to the micelle size. This is a fact frequently observed[12].

Finally, Equation (4) predicts that, when $[PdCl_4^{-2}]$ is small and $[H_2]$ (or $[NaBH_4]$) high enough, the initial reaction rate becomes independent of $[H_2]$ (or $[NaBH_4]$), $v_0 = k_1[S]_0[PdCl_4^{-2}]_0$. This predicted behavior was verified experimentally when we reduced the precursor concentration to $[Pd^{2+}]$ = 2 mM, as shown in Figure 8.

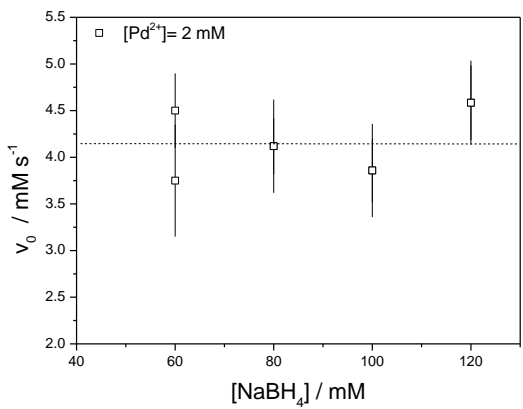

Figure 7: Initial reaction rate as a function of reducing agent concentration for an initial Palladium salt concentration of $[PdCl_4^{-2}]$= 2mM. Note that the reaction rate becomes constant for $[NaBH_4]$>60 mM, a behavior predicted for the model.

**Effect of Temperature on the Reaction Rates.**

Equation (11) also shows the dependence of the reaction rate on temperature via the coefficient of diffusion $D_{eff}$. If we assume an Arrhenius-like behavior, it should be, $D_{eff} \sim exp(-E_a/k_bT)$; thus, $ln(v_0) \sim 1/T$, if all the other parameters do not change with T. Because $D_{eff}$ includes the diffusion of $H_2$ through the micelle membrane, it also incorporates all the effects due to the specific characteristics of the micelle surfactant/co-surfactant monolayer; thus, it should depend on the chemical system surfactant/co-surfactant and also on the oil phase used in the formulation of the microemulsion. All these predictions could be verified experimentally. As an example, the effect of temperature will be analyzed below.

Note that if $r$ changes as T changes, its dependence on temperature, $r(T)$, must also be added into Eq. (11). Thus, we need to know first the dependence of the microemulsion micelle size on temperature, $r(T)$. Figure 9 shows the micelle diameter as measured by dynamic light scattering (hydrodynamic diameter) as a function of temperature for microemulsions containing only water, water + Pd salt at different concentrations, and water + $NaBH_4$ also at several concentrations, all prepared at the same molar ratio of water to DTAB ($W_0$=15). The size does not depend on reactant concentrations; it only depends on temperature. The micelle radius varies following a power law, $r = 1.37 + 6.65 (T - 273.15)^{-0.18}$.

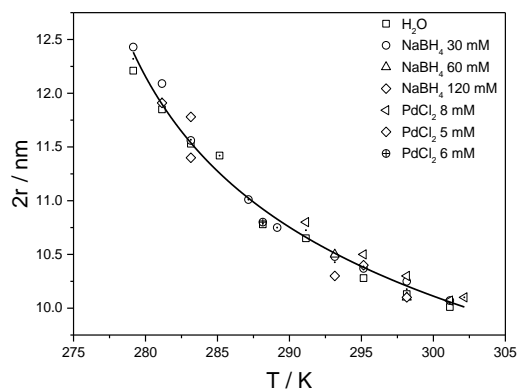

Figure 8: Hydrodynamic diameter (2r) of micelles as a function of temperature for microemulsions containing different amounts of reactants (before mixing and reacting). The line is a fitting curve with a power law: $r \sim (T-273.15)^{-018}$.

Introducing this dependence into Equation (11), and assuming an Arrhenius-like behavior for the effective diffusion coefficient, we can expect the following dependence of the initial reaction rate on temperature,

$$ln(v_0) \sim -\frac{E_a}{RT} + 3\ ln(1.37 + 6.65(T - 273.15)^{-0.18})\quad(12)$$

In Figure 8, we present results on the reaction rate as a function of temperature. Considering Equation 12, we fitted the experimental points with the following function, $ln(v_0) = A - \frac{E_a}{RT} + 3\ ln(1.37 + 6.65(T - 273.15)^{-a})$ ; from this fitting, we obtained $E_a$= 97 kJ and $a$ = 0.2. The last value is quite close to 0.18 obtained from light scattering experiments (shown in Figure 8); however, because $a$ is so small, an equally good (slightly less good) fit can be obtained by using just the Arrhenius-like equation, $ln(v_0) \sim -\frac{E_a}{RT}$. In this case, $E_a$ = 107 kJ/mol (25.5 Kcal/mol, which is a physically reasonable value[48] (see Figure 9).

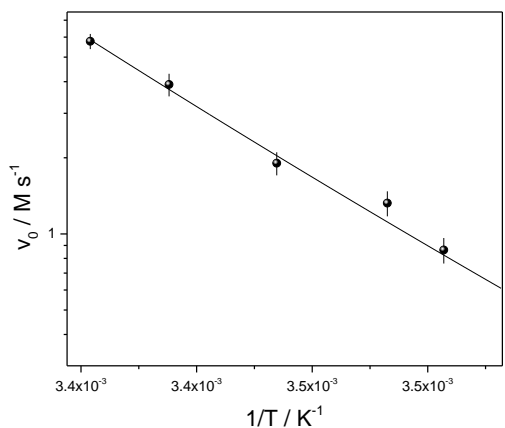

Figure 9: (a) initial reaction rate as a function of the inverse of temperature. The line is a fitting with an Arrhenius- like function.

**Nucleation, growth and size dependence.**

As it was mentioned in the Introduction, our hypothesis is that the particle size depends directly on the reaction rate (and indirectly on concentration, temperature, water-surfactant ratios, etc.). In Figure 10, we plotted the particle size measured by TEM microscopy as a function of the initial reaction rate, independently of the actual palladium precursor and reducing agent concentrations. A quite clear tendency can be observed: the lower the reaction rates, the larger the particle size and particle size polydispersity (Figure 11). Let us remark here that the size is correlated with the reaction rate and not with the actual reactant concentration (or temperature) we can get the same initial reaction rates for different relations of Pd salt and $NaBH_4$ concentrations (or different temperatures).

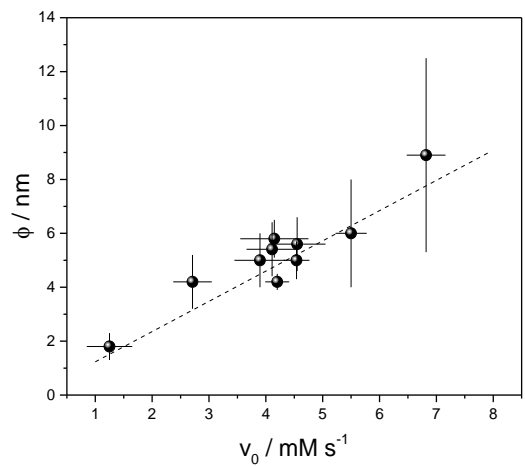

Figure 10: Particle size as a function of initial reaction rate. The line corresponds to a fitting: $\phi$(nm) ~ $v_0$ (mM/s).

The points and error bars in Figure 10 are the result of about a hundred of experiments on the reaction kinetics, such as those in Figure 1. In this figure, each point at a given measured value of the reaction rate includes experiments performed at different concentrations of precursor, reducing agents, and temperature given a similar initial reaction rate, $v_0$. Figure 10 shows that the particle diameter is approximately linear with the initial reaction rate, $\emptyset \sim v_0$; thus, from Eq. (11) and replacing [$H_2$] by [$NaBH_4$], the Pd particle sizes scale as,

$$\emptyset \sim v_0 \sim D_{eff}\left(\frac{V_w}{V_T}\right)^3 \frac{1}{r^3}[NaBH_4]\quad(13)$$

and the particle size will depend on the micelle size, water and total volume, the reducing agent concentration and the effective diffusion constant through the micelle surfactant monolayer. At risk of being repetitive, we stress again that, in Equation 13, the influence of the micelle size on particle sizes is not due to the confinement of the reaction within the walls of water pool of the micelles acting as a template; it is an indirect consequence of changes on the total diffusion area ($A_m$), the number of micelles (N), and the distance (l) between adjacent micelles, all of them a function of the micelle radii (r).

As mentioned before, the size prediction depending on water and total volume was recently reported for the synthesis of ($H_3O$)$Y_3F_{10}$ nanoparticles in microemulsions[23]. The effect of the concentration of the reducing agent on particle size predicted by our model was also observed in the synthesis of

Pd particles in reverse microemulsions using AOT as surfactant [17]. Equation 13 also predicts an increase of particle size with temperature, $D_{eff}$ increases, which produces an increase of the initial reaction rate, and then an increase in particle size. This predicted behavior with temperature was also observed in the production of Ag particles in reverse microemulsions using $NaBH_4$ as a reducing agent [50,51].

The dependency of particle size on the reaction rate was introduced via the empirical law extracted from the results of Figure 10. Thus, let us now discuss the process of nucleation and growth of the nanoparticles in reverse microemulsions in the context of nucleation and growth theories in solution [52,53]. One of the most well-known ideas from nucleogenesis theories is the explanation of the production of monodisperse colloids [54]. The classical nucleation theory predicts that the rate of nuclei production, $dN_u/dt$, where $N_u$ is the number of nucleus produced and t the time, yields [53]

$$\frac{dN_u}{dt} \sim e^{-\frac{\gamma^3 V^2}{(k_B T)^3 [ln(R)]^2}} \qquad (14)$$

being γ the interfacial tension, V the particle volume, R the supersaturation ratio, $R=S_{ss}/S_0$, $S_{ss}$ the actual concentration of atomic Pd, and $S_0$ the equilibrium solubility limit of atomic Pd ($S_{ss}>S_0$). This equation indicates that the rate at which the particle nuclei is produced increases with the supersaturation ratio, being this ratio the controlling parameter. In our case, this depends on the reaction rate; if it is slow, the production of a few nuclei reduces appreciably the Pd concentration and relieves the supersaturation, resulting in a reduction of the rate of nucleation (Eq. 14). Consequently, by reducing the reaction rate, the period in which nucleation can occur is reduced in such a way that monodispersed particles result from the uniform growth of the existing nuclei. In this regime, a balance between the rate of production of metallic Pd and the removal by diffusion onto the nuclei is attained. These monodisperse particles would be protected by the microemulsion micelles, which avoids coalescence among them.

On the contrary, when the rate of reaction is high, the rate of production of Pd metallic atoms during the chemical reaction becomes so rapid that the supersaturation remains, and the nuclei are produced continuously while the existing nuclei grow. In this regime, the size of any given particle will depend upon when it was created. As a consequence, the particle size distribution broadens, becoming polydisperse (see Figure 5). This explains the increased polydispersity, but to explain the increase in the particle size as the reaction rate increases, it is necessary to include in the discussion the dynamics of particle coalescence and Ostwald ripening (coarsening). The rapid production of nuclei simultaneously with the growth of the existing ones in conjunction with the processes of collision, fusion and intermicellar exchange of material should permit the coalescence of Pd particles and their growth by coarsening, a process that should not be possible if the production of Pd metallic atoms is slow. The previous discussion is just speculative and have to be verified experimentally.

## CONCLUSIONS

In this article, we presented a systematic study of the reaction kinetics of $PdCl_2$ with $NaBH_4$ for the production of metallic Pd nanoparticles in reverse microemulsions. Our work was based on the assumption that the reaction rate is the main factor controlling the final particle size. We choose this particular system to be studied because it allows us to follow the reaction rate by the consumption of one of the reactants, independently of the dynamics of the nanoparticle growth. On the basis of more than one hundred experiments, we found that the final particle size is well correlated with the initial reaction rate, and not directly correlated with parameters such as reactant concentrations or temperature. Those parameters effectively modify the final particle size, but indirectly by changing the reaction rates.

In order to explain the reaction kinetics in our systems, we departed from the commonly accepted view of considering the process of mixing the reactants by intermicellar exchange [9,10,45,55,56]: once the microemulsions containing the precursor and the reducing agent are mixed, the droplets collide, fuse, interchange the material and break apart again, being this process the unique way of producing the mixing of the reactants. In our system, we considered the possibility that the reducing agent diffuses out of the water pool into the continuous phase. This was based on three experimental facts: first, $NaBH_4$ in contact with water hydrolyzes to produce hydrogen; second, hydrogen is four times more soluble in benzene, the continuous phase, than in water; third, the capability of acting as a reducing agent of the $NaBH_4$ disappears after 2 hs, which indicates that the $H_2$ can effectively go out the water pools of the microemulsion.

Additionally, the reaction kinetics was experimentally found to be first order in NaBH$_4$ and zero order in the precursor of palladium. This last experimental fact leads us to consider the possibility of having a reaction mechanism mediated by the micellar interface, which provides sites for the condensation of the precursor ions PdCl$_4^{-2}$. The simple model that emerges from all these observations, if even naïve, allows to explain some of the experimental findings, such as the dependence on the reactant concentration, the observed reaction orders and the influence of temperature on the reaction rates. It also predicts how the reaction rates depend on the micelle size, the water volume and the microemulsion total volume. We think that this model has the merit of predicting quantitatively how the micelle size affects indirectly the size of the particles without the need of considering the water pools as close templates. Some of these predictions were indeed observed in the production of nanoparticles in reverse microemulsions[17,23,50,51], and some others need to be verified experimentally. In this respect, we will work in the near future on how the reaction kinetics depends on the micelle radii, the water content and the total microemulsion volume. Recall that the model predicts, for this particular system, $v_0 \sim r^{-3}$; $v_0 \sim V_w^3$ and $v_0 \sim V_T^{-3}$, respectively. The influence of the interfacial elasticity and the use of different oils as continuous phase, all parameters suspected of modifying the effective diffusion coefficients, will be studied as well. For that purpose, we will use different surfactants, co-surfactans and organic liquids in the formulation of the microemulsions but using the same model reaction, that is, the same reactants.

**Acknowledgments.** This work was supported, in part, by Universidad Nacional del Sur (UNS, Argentina) under grants PGI-UNS 24/F067 and PGI-UNS 24/F070; by Agencia Nacional de Promoción Científica y Tecnológica (ANPCyT, Argentina) under grants PICT-2013-2070 and PICT-2016-0787, and by Consejo Nacional de Investigaciones Científicas y Técnicas (CONICET, Argentina) under grants PIP-GI 2014 Nro 11220130100668CO.

The authors would like to thank Dr. Alberto Caneiro from Y-TEC (YPF Tecnología, Argentina) for his contribution to the HRTEM characterizations.

## REFERENCES.